\documentclass[american,authoryear, 12]{elsarticle}
\usepackage[LGR,T1]{fontenc}
\usepackage[latin9]{inputenc}
\usepackage{xcolor}
\usepackage{pdfcolmk}
\usepackage{textcomp}
\usepackage{mathrsfs}
\usepackage{graphicx}
\PassOptionsToPackage{normalem}{ulem}
\usepackage{ulem}

\makeatletter

\DeclareRobustCommand{\greektext}{%
  \fontencoding{LGR}\selectfont\def\encodingdefault{LGR}}
\DeclareRobustCommand{\textgreek}[1]{\leavevmode{\greektext #1}}
\ProvideTextCommand{\~}{LGR}[1]{\char126#1}

\providecommand{\tabularnewline}{\\}
\providecolor{lyxadded}{rgb}{0,0,1}
\providecolor{lyxdeleted}{rgb}{1,0,0}

\DeclareRobustCommand{\lyxsout}[1]{\ifx\\#1\else\sout{#1}\fi}

\usepackage{hyperref}

\makeatother

\usepackage{babel}
\begin{document}

\begin{frontmatter}{}

\title{Machine Learning for automatic identification of new minor species}

\author{Fr{\'e}d{\'e}ric Schmidt $^{1}$, Guillaume Cruz Mermy$^{1}$, Justin 
Erwin$^{2}$, S{\'e}verine Robert$^{2}$, Lori Neary$^{2}$, Ian R. Thomas$^{2}$, Frank Daerden$^{2}$, Bojan Ristic$^{2}$,
Manish R. Patel$^{3}$, Giancarlo Bellucci$^{4}$, Jose-Juan Lopez-Moreno$^{5}$, Ann-Carine Vandaele$^{2}$}

\address{$^{1}$ Universit{\'e} Paris-Saclay, CNRS, GEOPS, 91405, Orsay, France,
$^{2}$ Belgian Institute for Space Aeronomy (BIRA-IASB), Avenue Circulaire,
3 B-1180 Brussels Belgium,$^{3}$ School of Physical Sciences, The Open University, Milton Keynes, MK7 6AA, U.K.,$^{4}$ INAF-Istituto di Astrofsica e Planetologia Spaziali, Rome, ITALY,$^{5}$ Instituto de Astrof{\'i}sica de Andaluc{\'i}a CSIC}
\begin{abstract}

One of the main difficulties to analyze modern spectroscopic datasets is due to the large amount of data. For example, in atmospheric transmittance spectroscopy,
the solar occultation channel (SO) of the NOMAD instrument onboard the ESA ExoMars2016 satellite called Trace Gas Orbiter
(TGO) had produced $\sim$10 millions of spectra in $\sim$20000 acquisition sequences since the beginning of the mission in April 2018
until 15 January 2020. Other datasets are even larger with $\sim$billions of spectra for OMEGA onboard Mars Express or CRISM onboard Mars Reconnaissance Orbiter. Usually, new lines are discovered after a
long iterative process of model fitting and manual residual analysis.
Here we propose a new method based on unsupervised machine learning,
to automatically detect new minor species. Although precise quantification
is out of scope, this tool can also be used to quickly summarize
the dataset, by giving few endmembers ("source") and their abundances.

The methodology is the following: we proposed a way to approximate
the dataset non-linearity by a linear mixture of abundance and source spectra (endmembers). We used unsupervised source separation in form of non-negative
matrix factorization to estimate those quantities. Several methods are tested on synthetic and simulation data. Our approach is dedicated to detect minor species spectra rather than precisely quantifying them.
On synthetic example, this approach is able to detect chemical compounds
present in form of 100 hidden spectra out of $10^4$, at 1.5 times the noise level.
Results on simulated spectra of NOMAD-SO targeting CH$_{4}$ show that detection 
limits goes in the range of 100-500 ppt in favorable conditions. Results on real martian data from NOMAD-SO show that CO$_{2}$ and
H$_{2}$O are present, as expected, but CH$_{4}$ is absent. Nevertheless, we confirm 
a set of new unexpected lines in the database, attributed by ACS instrument Team to the {CO}$_{2}$ magnetic dipole.

\end{abstract}

\begin{keyword}
spectroscopy, atmosphere, data mining, machine learning, unsupervised,
source separation, non-negative matrix factorization
\end{keyword}

\end{frontmatter}{}


\section{Introduction}

In modern exploration science, one has to face a major challenge : how
to learn something new from analyzing a large dataset collection while taking into account what we already know. If the current knowledge overrides the analysis, the discovery of new elements may be difficult. Usually, in the field of spectroscopy,
one can compare laboratory spectra, model and observation spectra.
Going back and forth leads to discovery of new lines by identifying
unexpected residuals in the observation data (not expected by the
model). Sometimes, initial identification of lines can be wrong. As
an example, spectroscopic evidence of atmospheric CO$_{2}$ ice cloud
was reported after the discovery of an emission spike at a wavelength of 4.3 \textgreek{m}m from Mariner 6 and 7 infrared probings of the
bright martian limb \citep{Herr1970}, but this spectral feature was
mistaken for a resonant scattering band of CO$_{2}$ fluorescence
\citep{Lopez-Valverde_AnalysisnonLTE_PaSS2005}. 

For one single spectrum, one can use simulation algorithm (see for instance \citealp{Faisal_Systematicinvestigationshyperfine_JoQSaRT2020}). For large datasets, simplest ideas would be to scrutinize average spectra, or potential
band depth distribution. Unfortunately, in the case of low
signal-to-noise ratio (SNR, defined as signal / standard deviation of noise), such methods fail (as will be
illustrated in the toy example). Analyzing residuals after modeling is a good method but it requires a lot of work.

Several statistical tools with various approaches have been proposed, such as the Principal Component Analysis (PCA) \citep{Penttilae_Laboratoryspectroscopymeteorite_JoQSaRT2018, Geminale_Removalatmosphericfeatures_I2015}, or Independent Component Analysis (ICA) \citep{Shashilov_Latentvariableanalysis_JoQSaRT2006, Erard_ICAvirtis_JGR2009}, but most of them require a human operator to pick endmembers and trends since those methods are nothing more than a change of representation. Furthermore, none of these methods guarantees positivity of the \textit{component} (which are sometimes also called \textit{source}), which can be problematic during the interpretation. Recently, advanced machine learning methods based on non-negative matrix factorization have been proposed \citep{Lee_NMF_Nature1999, Moussaoui_JADE-BPSS_Neurocomp2008, Dobigeon_BPSS2_SignalProcessing2009, Schmidt_ImplementationBPSS_TGRS2010, Gillis_AcceleratedMultiplicativeUpdates_NC2012, Hinrich_ProbabilisticSparseNon_Chapter2018}. This approach is completely different from PCA/ICA: each source is positive and represents an endmember / a trend. A source is not one spectrum extracted from the dataset but a statistical reconstruction. By using this approach, the human operator doesn't have to identify endmembers/trends anymore, since they are automatically picked by the algorithm in form of source. Furthermore when there are statistical / spectral correlations between sources PCA/ICA fails because it assumes orthogonality / independence, which is not the case for non-negative matrix factorization.

Based on this new approach, we propose a tool:
\begin{itemize}
\item to give an overview and quickly summarize a large and complex spectroscopic dataset with simple variables
\item to detect potential new spectroscopic features (unexpected minor species, new absorption lines,...)
\item to be performed in a fully blind way (without prior information on neither the spectra,
nor the abundances).
\end{itemize}

The target observation type of this study is solar occultation. This measurement principle has been proposed as early as 1900,
an interesting review was published by \citet{Smith_Studyplanetaryatmospheres_RoG1990}.
Several recent instruments used this technique to investigate the composition of the Earth's
(SCIAMACHY/ENVISAT \citealt{Bovensmann_SCIAMACHYMissionObjectives_JotAS1999}),
Mars' (SPICAM \citealp{Bertaux_SPICAM_PaSS2000}) or Venus' atmospheres
(SPICAV \citealp{Bertaux_SPICAVVenusExpress_PaSS2007}). Here we will
focus on the recent NOMAD instrument \citep{Vandaele_NOMAD_PaSS2015},
and especially the SO channel, designed to study the Martian atmosphere
and its trace gases, such as methane. Indeed the presence
of CH$_{4}$ on Mars is a very hot topic for the planetary science community
\citep{Giuranna_Independentconfirmationmethane_NG2019,Korablev_Nodetectionmethane_N2019,Moores_Methaneseasonalcycle_NG2019}.
In the present article, we propose to apply the tool for potential
CH$_{4}$ detection. Nevertheless, the approach can be extended to
other types of spectroscopic measurements.

\section{Dataset\label{sec:Dataset}}

We propose here to focus on the Nadir and Occultation for MArs Discovery
(NOMAD) instrument onboard ESA's ExoMars Trace Gas Orbiter and especially the Solar Occultation (SO) channel
\citep{Vandaele_NOMAD_PaSS2015}. NOMAD is a compact, high-resolution,
dual channel IR spectrometer (SO and LNO) coupled with a highly miniaturized
UV-visible spectrometer (UVIS), capable of operating in different
observation modes: solar occultation, nadir and limb. 

The SO channel operates at wavenumbers from 2320 cm$^{-1}$ to 4550 cm$^{-1}$
(wavelength 2.2 to 4.3 $\mu$m), using an echelle
grating with a groove density of 4 lines/mm in a Littrow configuration
in combination with an Acousto-Optic Tunable Filter (AOTF) for spectral
order selection. The width of the selected spectral ranges is recorded by 320 spectels (spectral element) and varies
from 20 to 35 cm$^{-1}$ depending on the selected diffraction order.
The detector is an actively cooled HgCdTe Focal Plane Array. SO achieves
an instrument line profile resolution of 0.15 cm$^{-1}$, corresponding
to a resolving power \textgreek{l}/\textgreek{Dl} of approximately
25000. All details of the instrument are available in \citealt{Neefs_NOMADspectrometeron_AO2015} and  \citealt{Vandaele_NOMADanIntegrated_SSR2018}.
The orders with the maximum sensitivity to CH$_{4}$ are: 119, 134
and 136. We will use the data from the beginning of the mission in April
2018 until 15 January 2020, in calibration version 1p0a. Due to temperature change, the spectral registration varies, producing a shift up to $\sim$10 spectels. We corrected it by aligning the full dataset to a reference spectra (arbitrarily choosen with the maximum band depth of water) by cross-correlation. No sub-spectel resampling has been performed but a simple shift. When the calibration will be improved, this step will most probably be replaced by a routine correction.
The data
are available on the ESA/Planetary Science Archive after a 6 months
embargo period.

\section{Method}

In this section, we first describe the data pretreatment required
for non-negative matrix factorization purpose followed by the data
mining method.

\subsection{Data pretreatment \label{subsec:Data-pretreatment}}

After calibration, the NOMAD SO spectra are in transmittance $T=I/I_{0}$,
depending on wavenumber $\nu$, with $I$ the observed light intensity
trough the atmosphere and $I_{0}$ the solar spectra measured outside the atmosphere.

Assuming that the atmosphere is homogeneous, and that multiple scattering
and refraction are negligible \citep{Smith_Studyplanetaryatmospheres_RoG1990,Bovensmann_SCIAMACHYMissionObjectives_JotAS1999},
the optical depth $\tau$ is a linear combination of $E(\nu)$
the total extinction, and $\epsilon$ the slant column density, for
each chemical species $i$:
\begin{equation}
\tau(\nu)=-\log T(\nu)\approx\sum_{i=1}^{N_{S}}E_{i}(\nu).\epsilon_{i}+MC(\nu)\label{eq:optical_depth}
\end{equation}

with $N_{S}$, the total number of species and $MC(\nu)$ a modeled continuum described below.

The slant column density $\epsilon$ is directly related to the total number of
particles $N(s)$ along the line of sight $s$:
\begin{equation}
\epsilon=\int N(s)ds\label{eq:slant_column}
\end{equation}

While the extinction by gas is usually
highly structured, absorption by particles, scattering by molecules
and particles, and also reflection at the surface are broadband features. Such large features are modeled by a continuum $MC(\nu)$, often taken as a polynomial, 
that is filtered out. 

The problem with this continuum removal rationale is that when the optical depth is
large, the SNR is decreased and the noise effect on continuum removal amplified (see Sup. Mat.).

Instead of using this rationale, we propose to first correct for the
continuum $C(\nu)$ in the transmittance space:

\begin{equation}
T^{*}(\nu)=T(\nu)-C(\nu)\label{eq:continuum-removal}
\end{equation}

Then convert the spectra into absorbance:

\begin{equation}
X(\nu)=1-T^{*}(\nu)\label{eq:}
\end{equation}

The final step is the linear mixture :

\begin{equation}
X(\nu)\approx\sum_{i=1}^{N_{S}}S_{i}(\nu).A_{i}\label{eq:LinearMixture}
\end{equation}

with $S_{i}(\nu)$ the source spectra and
$A_{i}$ the spectral abundance. In this description, the
physical meaning of $S_{i}(\nu)$ and $A_{i}$ is lost but the apparent SNR is dramatically increased, which is much more important for our analysis.
Nevertheless, assumptions required in eq. \ref{eq:optical_depth}
are usually not relevant. Radiative transfer model used for precise
quantification is highly non-linear.

One has to consider that this unsupervised linear unmixing problem is already very difficult for machine learning. Solving non-linear model in a unsupervised way is a research area that is clearly not solved yet. In addition, we would like to focus on spectral detection, rather than quantification. Thus, we will focus on $S(\nu)$ much more than $A$. We will show that for linear, but also non-linear simulation and real data, meaningful $S(\nu)$ can be retrieved. Due to non-linearity, $A$ may differ significantly from truth, but the big tendencies should be respected. 
After the quick-look analysis, estimating $S_{i}$ and $A$, one must go back to the real data. The most trivial strategy is to pick the spectra $X$ out of the collection, with the highest abundance of a selected source $S_{i}$.

In the following, we will use the continuum estimation $C(\nu)$
using asymmetric least square \citep{Eilers_Baselinecorrectionasymmetric_LUMCR2005},
with parameters : $\nu_{smooth}=10^{3}$ and $p=1-10^{-2}$, 10
number of iterations.

\subsection{Non negative matrix factorization \label{subsec:NMF-model}}

For a collection of spectra, eq. \ref{eq:LinearMixture} can be written
in matrix form $\mathbf{X}_{kj}\mathbf{\approx}\mathbf{S}_{ki}.\mathbf{A}_{ij}$,
with $i$ the source index (from 1 to $N_{S}$) , $j$ the observation
index (from 1 to $N_{O}$) and $k$ the wavenumber index (from 1
to $N_{\nu}$). Thus, one have to estimate $\mathbf{S}$ and $\mathbf{A}$,
by minimizing the objective function:

\begin{equation}
F=\left\Vert \mathbf{X-S.A}\right\Vert ^{2}\label{eq:NNMF}
\end{equation}

with $\left\Vert .\right\Vert $, the Frobenius norm (usual $L_{2}$
norm).

Several algorithms have been proposed to solve this problem, subject
to positivity (both $\mathbf{S}$ and $\mathbf{A}$ are non-negative). Such problem is called Non negative Matrix Factorization (NMF). This constraint is important to keep the physical meaning, but also
to promote sparsity of $\mathbf{S}$ (a signal is sparse when most of the values are close to zero except several non-zero values). Let $\dot{\mathbf{S}}$
and $\dot{\mathbf{A}}$ be the estimation of those quantities.

\paragraph{MU} We propose to use the Multiplicative Updates (MU) of \citet{Lee_NMF_Nature1999}
accelerated by \citet{Gillis_AcceleratedMultiplicativeUpdates_NC2012}.
We used the convergence parameter $\alpha_{MU}=1$. Other alternative
algorithms are possible but give equivalent results since they minimize the same cost function. This algorithm has the advantage of very fast computation time but the result may depend on initialization.

\paragraph{BPSS2} We propose to test another kind of algorithm: the Bayesian Prior Source
Separation \citep{Moussaoui_BPSS_IEEE2006, Dobigeon_BPSS2_SignalProcessing2009}, that has
been optimized \citep{Schmidt_ImplementationBPSS_TGRS2010}, hereafter called BPSS2. This
algorithm has the main advantage to account for extra constraint
: the sum-to-one or sum-lower-than-one on the abundances ($\sum_{i}A_{ij}=1$) that also
promotes sparsity of $\mathbf{S}$. This algorithm, based on Monte
Carlo approach is much more time consuming. One approach to reduce the computation time is to select only relevant spectra out of the dataset \citep{Moussaoui_JADE-BPSS_Neurocomp2008}, but then the statistics may be biased \citep{Schmidt_ImplementationBPSS_TGRS2010}. Thanks to the advances of computer capabilities, we propose to treat the full dataset. This kind of algorithm is very slow but since the formulation is Bayesian, it converge toward an unique solution.

\paragraph{psNMF} In order to regularize the problem of eq. \ref{eq:NNMF}, one can
add an extra penalization term to enforce sparsity on $\mathbf{A}$
(only few non zeros elements in $\mathbf{A}$) \citep{Kim_Sparsenonnegative_B2007}
:

\begin{equation}
F=\left\Vert \mathbf{X-S.A}\right\Vert ^{2}+\lambda\left\Vert \mathbf{A}\right\Vert _{1}\label{eq:sparseNNMF}
\end{equation}

With $\left\Vert .\right\Vert _{1}$, the $L_{1}$ norm. The first term is called data attachment term (the usual squared difference). The second is called regularization term. The problem
with this approach, is that hyperparameter $\lambda$ is not known
and has to be tuned manually. A recent approach has been proposed
to solve this problem in the Bayesian framework \citep{Hinrich_ProbabilisticSparseNon_Chapter2018}.
The main idea is to encompass all variables and hyperparameters in
a unique problem that is estimated with variational update principle.
We will refer this algorithm to probability sparse NMF (psNMF). This algorithm has the advantage to have a reduced computation time and no hyperparameter tuning. It also has a regularization term to avoid strong dependence of the initialization on the final solution.

In order to estimate the precision of the reconstruction, we used the
Root Mean Square Difference $RMSD$:

\begin{equation}
RMSD=\frac{\sqrt{\left\langle \left(\mathbf{X-\dot{S}.\dot{A}}\right)^{2}\right\rangle }}{\left\langle \mathbf{X}\right\rangle }\label{eq:RMSD}
\end{equation}

With $\left\langle .\right\rangle $, the mean.

Once the sources are estimated, we quantify their relevance for
the global dataset. From the total reconstruction $\mathbf{\dot{X}}_{kj}\mathbf{=}\mathbf{\dot{S}}_{ki}.\mathbf{\dot{A}}_{ij}$, for all $i$,
we can estimate the contribution of source $i'$, that is to say:
$\mathbf{\dot{X}}_{kj}^{i}\mathbf{=}\mathbf{\dot{S}}_{ki'}.\mathbf{\dot{A}}_{i'j}$.
Thus, the relevance of source $i$ is defined as:

\begin{equation}
R^{i}=\frac{\left\langle \left|\mathbf{\dot{X}}^{i}-\mathbf{\dot{X}}\right|\right\rangle }{\left\langle \mathbf{\dot{X}}\right\rangle }\label{eq:Relevance}
\end{equation}

This definition is convenient since the sum of all $R^{i}$ is one (this property is only present when sources and abundances are positive)
and we can easily estimate the \% contribution of each source in
the final reconstruction. One has to note that relevance is not a measure of presence or not of a minor specie (for instance CH$_{4}$) but a measure of how important is the source over the dataset. Major species,  should always have a larger relevance than minor species. In the following, we plot all sources results by decreasing order of relevance.

\subsection{Band depth (BD)} \label{sec:band_depth}

We used the following band depth definition, difference of the geometric
mean of two reference wavenumbers in the continuum, compared to the
band:

\begin{equation}
BD=X(\nu_{l})^{\frac{\nu_{c}-\nu_{l}}{\nu_{r}-\nu_{l}}}.X(\nu_{r})^{\frac{\nu_{r}-\nu_{c}}{\nu_{r}-\nu_{l}}}-X(\nu_{c})\label{eq:band_depth}
\end{equation}

with $X$ the observed spectra in transmittance, $\nu_{c}$ the
wavenumber of the center of band, $\nu_{l}$ the wavenumber of
the reference level on the left (smaller wavenumber), $\nu_{r}$ the
wavenumber of the reference level on the right (larger wavenumber).

\section{Synthetic tests\label{sec:synthetic-tests}}

We simulated several synthetic observations in different conditions, to mimic the case of NOMAD-SO. The first section describes a simple toy model example and the second one presents extensive tests of this toy model with various cases. By \textit{hidden spectra}, \textit{hidden compounds} and \textit{hidden CH$_{4}$}, we always refer to a spectral dataset with a dominant major component (here water) and a minor specie (here CH$_{4}$). The goal of the proposed approach is to pick up a source, containing CH$_{4}$ only.

\subsection{Toy example\label{subsec:Toy-example}}

\subsubsection{Synthetic dataset \label{subsec:toy_model_synthetic_dataset}}

In order to demonstrate the usefulness of our method, we propose here
a toy example in a very difficult case. We will see that usual method
fails detection but our method is able to detect the hidden compounds.

\begin{figure}
\hfill{}\includegraphics[viewport=50bp 200bp 530bp 620bp,clip,width=1\columnwidth]{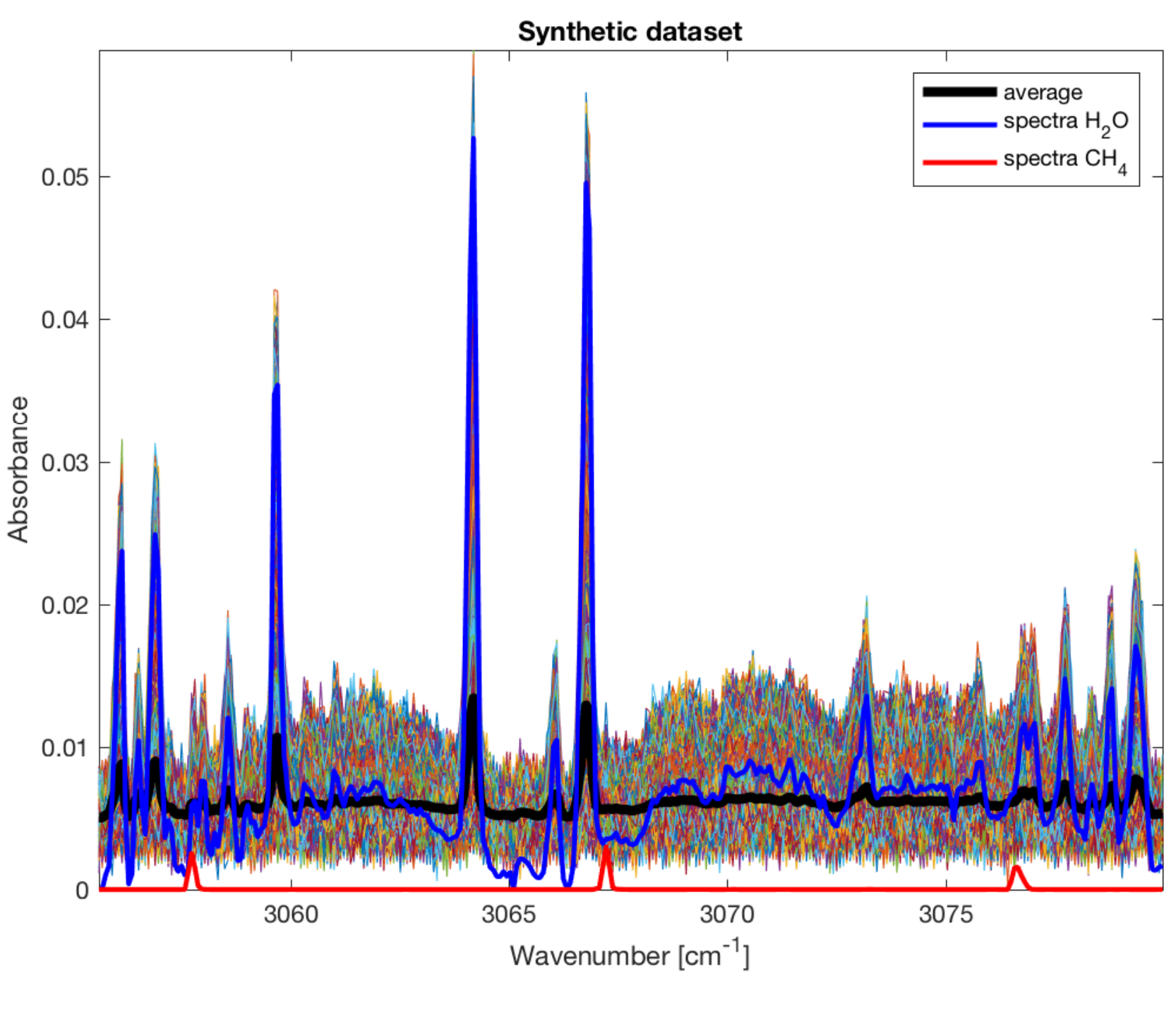}\hfill{}

\caption{Synthetic dataset containing $10^{4}$ spectra with various abundances of H$_{2}$O and 
100 containing CH$_{4}$ at 3-$\sigma$ level of the noise. In blue the reference spectra $S_{H_{2}O}$ of H$_{2}$O (coming from actual data analysis). In red the reference spectra $S_{CH_{4}}$ of CH$_{4}$ (from theoretical data).
\label{fig:Example-of-one-worst-spectra}}
\end{figure}

For this toy example, we simulate a linear mixture of $N_{O}=10^{4}$ observations spanned over $N_{\nu}=320$ spectels (see fig. \ref{fig:Example-of-one-worst-spectra})
similar to order 136 of NOMAD-SO. Each spectrum is a mixture of a spectra
of water vapor $S_{H_{2}O}$ (coming from one actual source estimated
from real data using psNMF) and theoretical methane $S_{CH_{4}}$ from \citet{Villanueva_PlanetarySpectrumGenerator_JoQSaRT2018},
with corresponding abundances $A_{H_{2}O}$, $A_{CH_{4}}$:

\begin{equation}
X=S_{H_{2}O}.A_{H_{2}O}+S_{CH_{4}}.A_{CH_{4}}+n\label{eq:toy_example}
\end{equation}

\begin{figure}

\hfill{}\includegraphics[width=1.0\columnwidth]{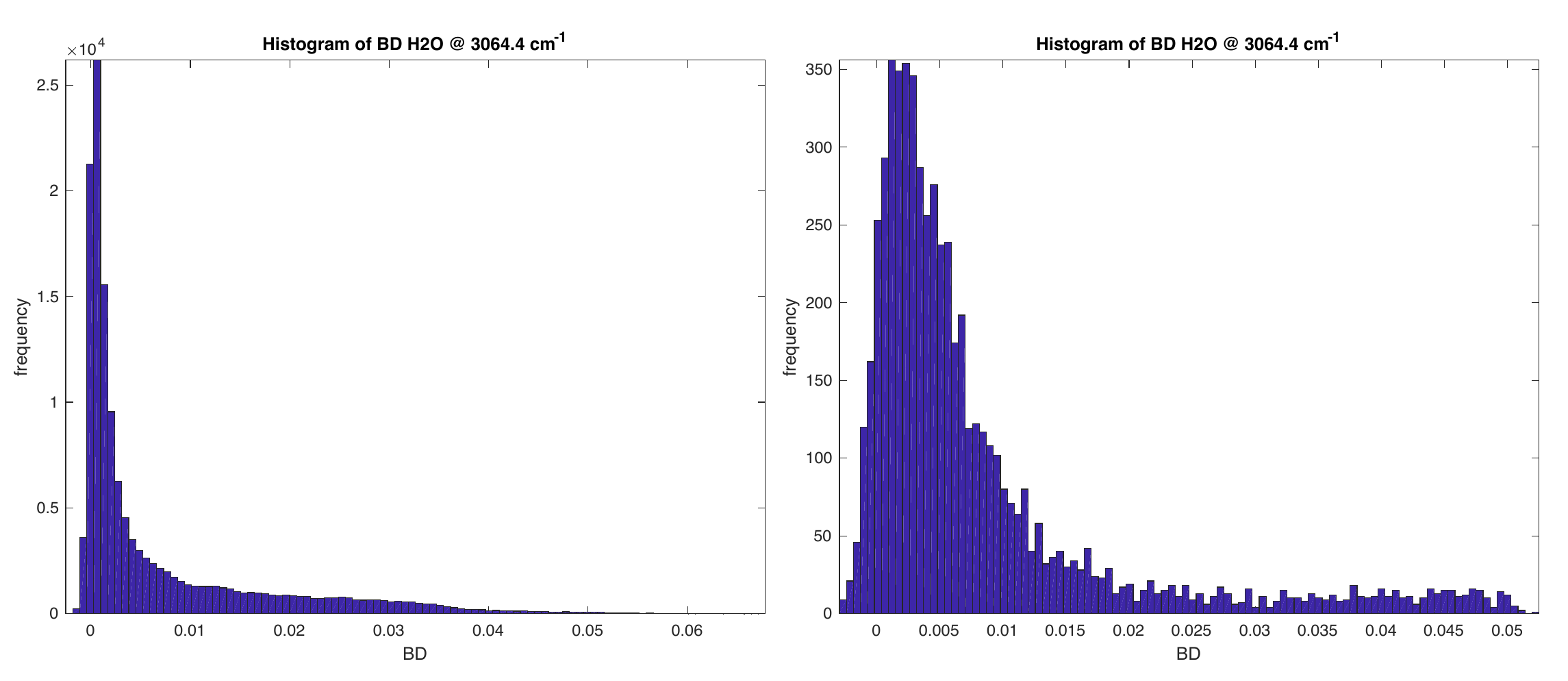}

\caption{Water vapor Band Depth distribution (left) in the real observation
(right) modeled by the toy example. \label{fig:Example-of-one-worst-distributionBD_H2O}}
\end{figure}

The noise $n$ is assumed to be a Gaussian process with a standard deviation of $\sigma$=0.001 and no bias: $n=\mathscr{G}(0,\sigma)$. All spectra contain pure water vapor with a coefficient following $A_{H_{2}O}=5/6.\mathcal{\beta}(1,10)+1/6.\mathcal{U}(0,1)$,
a mixture of beta ($\mathcal{\beta}$) distribution for 5/6 of the sample and an uniform ($\mathcal{U}$) distribution for 1/6 of the sample. This process mimics well the water vapor band depth distribution (BD, see definition in section \ref{sec:band_depth}) of the real dataset (see Fig.\ref{fig:Example-of-one-worst-distributionBD_H2O}).
As the baseline of $S_{H_{2}O}$ is not zero, we also mimic baseline correction errors. In addition 100 spectra out of 10000 contain methane with $A_{CH_{4}}=1$, such that the band depth of $S_{CH_{4}}$ is at 3-$\sigma$ level. Please note that the model to generate the data is not fulfilling the sum-to-one constraint, but fully fulfilling the positivity constraint. Given the defined noise and signal level, the $RMSD$ expected for a perfect reconstruction of the signal (and not the noise) is 0.16.

The final synthetic dataset is represented in Fig. \ref{fig:Example-of-one-worst-spectra}. 

In order to check the quality of the estimation, we simply compute the correlation coefficient between $S_{CH_{4}}$ and the estimated $N_{S}$ sources $\dot{\mathbf{S}}$, using: 

\begin{equation}
Q = corr\left\{ S_{CH_{4}},\mathbf{\dot{S}_{:i}}\right\} \label{eq:correlation_CH4spectra}
\end{equation}

The $i$th source with the maximum correlation is identified to $CH_{4}$ contribution. The value to the maximum correlation is used as metric to assess the quality of the retrieval.

\subsubsection{Results \label{subsec:toy_model_example}}

By plotting the 10000 samples of the dataset, one is able to identify easily the H$_{2}$O bands. Nevertheless, we cannot observe the target CH$_{4}$ in the average spectrum, even at 3-$\sigma$ level, because it is lost in the baseline changes. 

\begin{figure}

\hfill{}\includegraphics[width=1.0\columnwidth]{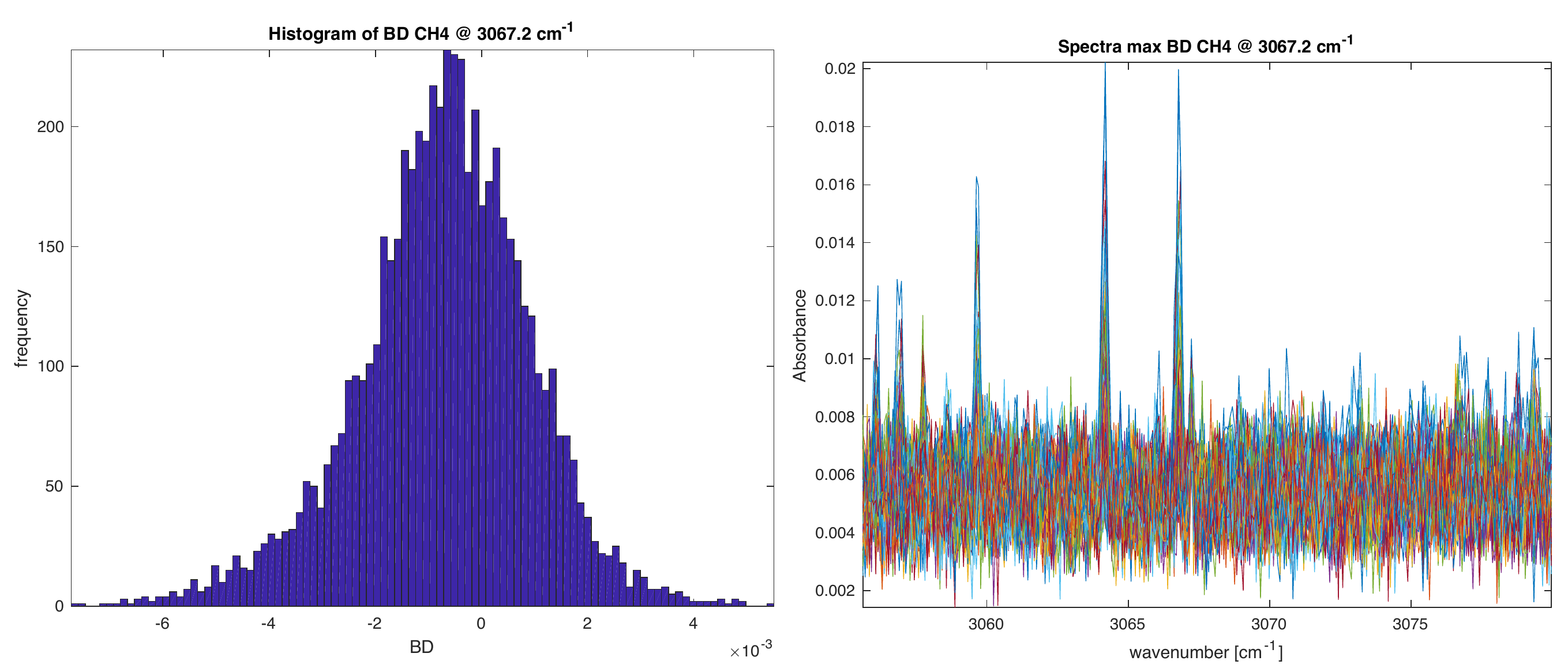}

\caption{(left) Histogram of Band Depth at 3067.2 cm$^{-1}$ from the dataset containing
100 CH$_{4}$ at 3-$\sigma$ level out of $10^{4}$ spectra. (right) 100
spectra with the maximum Band Depth at 3067.2 cm$^{-1}$ specific of CH$_{4}$. Signal is dominated by water and by noise. No specific signature of CH$_{4}$ is visible. \label{fig:Example-of-one-worst-BD_CH4}}
\end{figure}

The second simple tool for detection would be the analysis of the
band depth. Figure \ref{fig:Example-of-one-worst-BD_CH4} (left) shows the histogram of the main CH$_{4}$ band that exhibits no sign of the presence of CH$_{4}$ (no asymmetry in the positive part). Figure \ref{fig:Example-of-one-worst-BD_CH4} (right) represents the 100 spectra with the maximum CH$_{4}$ BD at 3067.2 cm$^{-1}$. Again, no particular elements can be used to argue for detection.

\begin{figure}

\hfill{}\includegraphics[viewport=45bp 220bp 530bp 620bp,clip,width=1\columnwidth]{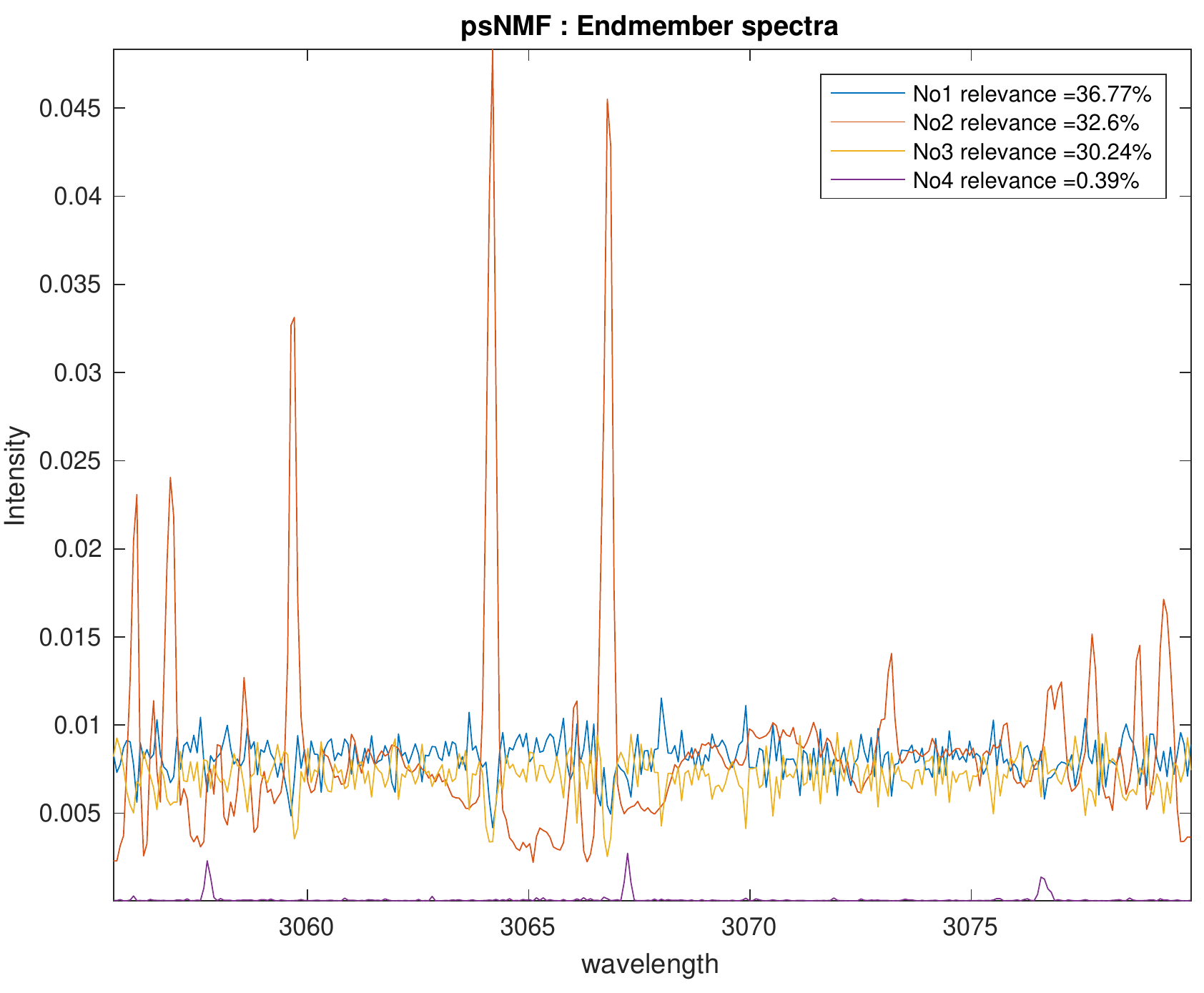}\hfill{}

\caption{Results of the psNMF algorithm for $N_{S}=4$. Sources 1 and 3 are
identified to the level with significant noise contribution, source
2 is identified to H$_{2}$O (correlation coef. with groundtruth 0.99), and source 4 is CH$_{4}$ (correlation coef. with groundtruth 0.98). Relevance
is computed from Eq. \ref{eq:Relevance}. \label{fig:Example-of-one-worst-result-MU}}
\end{figure}

Figure \ref{fig:Example-of-one-worst-result-MU} represents the results from the non-negative matrix factorization using psNMF algorithm. One can clearly identify both H$_{2}$O and CH$_{4}$ sources. Since those 2 chemical compounds are not correlated in abundance, ($A_{H_{2}O}$ and $A_{CH_{4}}$ are independent), two different source spectra are identified. Please note that the relevance of source 4 is very low (0.4\%), meaning that only 0.4\% of the variability in the dataset is due to CH$_{4}$, a very low value, as expected for minor species. 

In this case, the correlation coefficient between estimated abundances $\mathbf{\dot{A}_{4:}}$ and true ones $A_{CH_{4}}$ is 0.73. Since the quantification of abundance is a more difficult problem, we will not pay excessive attention on this parameter.

\subsubsection{Convergence and computation time \label{subsec:Convergence}}

We set the MU algorithm convergence to relative difference of the cost function $< 10^{-8}$ and a maximum running time of 1000 seconds. For psNMF, we set the relative difference of the cost function to $< 10^{-7}$ and a maximum iteration to 2000. For BPSS2, we compute a minimum burn in of 1000 iterations and after that when the long term statistics (1000 last iterations) of the Markov Chain is close to the short term statistics (100 last iterations), convergence is considered to be reached. Then another 1000 iterations are computed to estimate the final solution statistics.

We run the 3 identified tools 10 times on the
same dataset with different noise realization, and compute mean and standard deviation from these 10 experiments. Results are presented in Table \ref{tab:Convergence-and-computation-time}. One can clearly see that the even if the convergence is set, there is a high variability in MU results, due to the lack of regularization.
On this particular example, the best is clearly psNMF algorithm.

The $RMSD$ is computed for all cases and shown in Table \ref{tab:Convergence-and-computation-time}. We can observe that the value is almost equivalent, around 0.146, for all method but MU is slightly better, due to the fact that the cost function has no other term. MU algorithm is just minimizing the reconstruction. As a comparison, the $RMSD$ expected for a perfect reconstruction of the signal (and not the noise) of this toy example is 0.16. With 5 sources (significantly more than the 3 sources defined in this toy example), noise is also encompassed within the approximated linear model, as expected.

The quality $Q$ is the only parameter to assess the quality of the algorithm to detect minor specie (here CH$_{4}$). In this particular toy example, psNMF seems to be the best algorithm, providing a source correlated with groundtruth CH$_{4}$ with a correlation coefficient up to 0.8. We will extensively test this performance in the next section.

We also estimate the computation time on a 2.9 GHz Intel Core i7 with 16 Go DDR3 RAM as an example. All algorithms are implemented in  \textcopyright Matlab using parallelized matrix computation. Results, presented in Table \ref{tab:Convergence-and-computation-time}, demonstrate that MU is faster than psNMF but both are clearly less resources consuming than BPSS2. From the computation time and efficiency, we excluded BPSS2 from the next tests.

\begin{table}
\begin{tabular}{|c|c|c|c|}
\hline 
 & MU & psNMF & BPSS2\tabularnewline
\hline 
\hline 
Quality $Q$& 0.35$\pm$0.12 & 0.822 $\pm$0.005 & 0.41$\pm$0.06\tabularnewline
\hline 
$RMSD$ relative error & 0.1455$\pm$$2.10^{-6}$ & 0.1461$\pm$$5.10^{-6}$ & 0.1468$\pm$$3.10^{-4}$\tabularnewline
\hline 
Computation time (s) & 13$\pm$8 & 46$\pm$9 & 413$\pm$21\tabularnewline
\hline 
\end{tabular}

\caption{Results (mean and standard deviation) from 10 realizations of a toy synthetic example with $N_{S}=5$ (in agreement with next section on synthetic tests), $N_{O}=10000$, $N_{\nu}=320$ and 300 CH$_{4}$ spectra hidden at a level of 1 std of the noise. Quality is computed as a correlation coefficient (see Eq. \ref{eq:correlation_CH4spectra}).
$RMSD$ is computed from Eq. \ref{eq:RMSD}. Computation time is expressed in second. \label{tab:Convergence-and-computation-time}}
\end{table}

\subsection{Extended synthetic tests\label{subsec:Noise-level}}

\begin{figure}

\hfill{}\includegraphics[width=1.0\columnwidth]{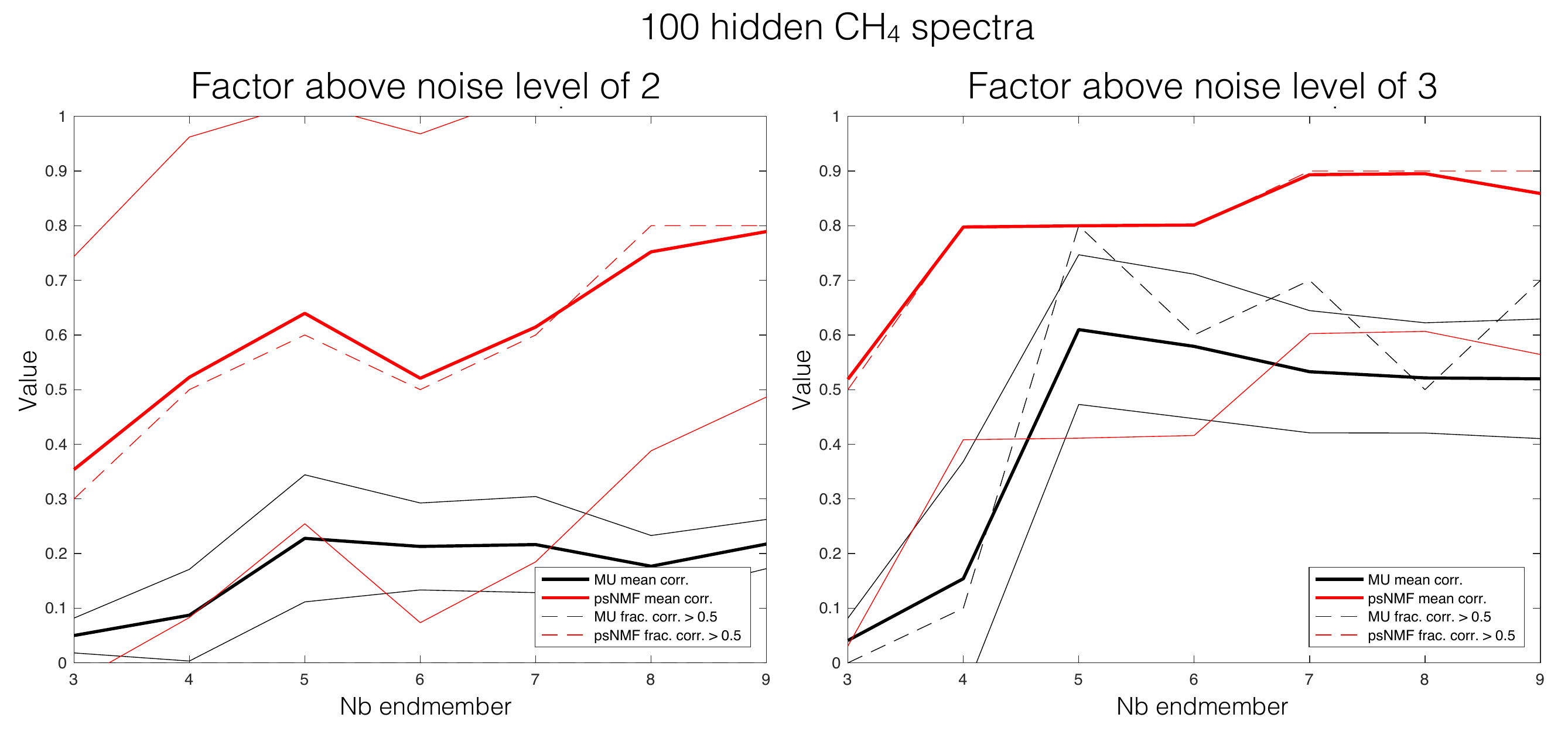}\hfill{}

\caption{Results of the MU and psNMF algorithm for $N_{S}=3$ to $9$, $N_{O}=10000$,
$N_{\nu}=320$, as a function of the number of source. The average $Q$ of 10 realizations of the best estimated
source (thick lines and standard deviation in thin lines) and the fraction
of acceptable results (with $Q>0.5$). (left) with a factor above noise level of 2 (right) with factor above noise level of 3. \label{fig:results_Ns}}
\end{figure}

For the first set of tests, we used the same toy model described in section \ref{subsec:Toy-example}, except with 100 CH$_{4}$ spectra hidden at a level of 2 and 3 standard deviation of the noise (this number is called ``factor above noise level''). In order to have robust results, we made 10 realizations and averaged the results.

Figure \ref{fig:results_Ns} represents the results as a function of the number of sources $N_{S}$.  It presents two quality indicators of the results: the average correlation coefficient $Q$ (see Eq. \ref{eq:correlation_CH4spectra}) and the fraction of realization with acceptable results (with $Q>0.5$). We can observe that the psNMF is always better than MU on average at cost of an higher variability (higher standard deviation). Adding sources seems to always increase the detection until reaching a plateau around $N_{S}=5$. Adding more sources will not drastically increase/decrease the source estimation. Nevertheless, it requires more computation time for a larger number of source ( approximately x2 between 3 and 9 sources but the computation time always stays below 200 seconds). 

For the second set of tests, we used the same toy model, except with 50 and 100 CH$_{4}$ spectra hidden at a level of 0.7, 1, 1.2, 1.5, 2.0, 2.5 and 3 standard deviation of the noise (this number is called ``factor above noise level''). In order to have robust results, we made 10 realizations and averaged the results. Results are always with $RMSD<0.18$ with an average $\sim 0.16$. $RMSD$ from the noise level is 0.16 whatever the experiment (the CH$_{4}$ is low enough so that it's contribution to $RMSD$ is negligible), so the reconstruction is in average as expected.

\begin{figure}

\hfill{}\includegraphics[width=1.0\columnwidth]{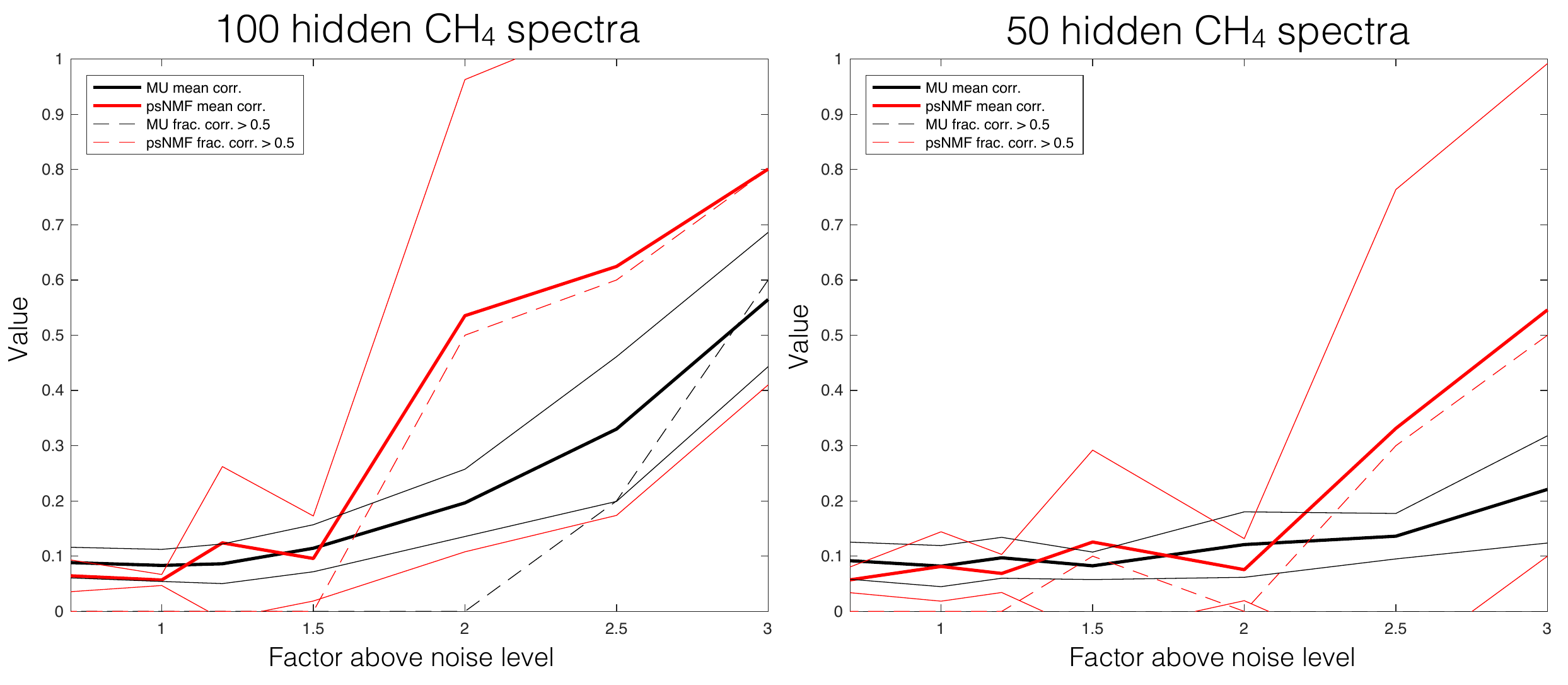}\hfill{}

\caption{Results of the MU and psNMF algorithm for $N_{S}=5$, $N_{O}=10000$,
$N_{\nu}=320$, as a function of the factor above noise level. The average $Q$ of 10 realizations of the best estimated
source (thick lines and standard deviation in thin lines) and the fraction
of acceptable results (with $Q>0.5$). (left) with 100 hidden CH$_{4}$ spectra
(right) with 50 hidden CH$_{4}$ spectra \label{fig:results_noise_level}}
\end{figure}

Figure \ref{fig:results_noise_level} presents two quality indicators of the results: the average correlation coefficient $Q$ (see Eq. \ref{eq:correlation_CH4spectra}) and the fraction of realization with acceptable results (with $Q>0.5$). Both indicators indicate that the method psNMF clearly outperforms MU at high factor above noise level. 
From our visual inspection of the results, we define the detection limit when at least 50\% of the results are with $Q>0.5$ (correlation coefficient $>$ 0.5). This definition is debatable but there is no absolute way of defining it. Figure \ref{fig:results_noise_level} shows that the detection limit is at 1.5 factor above noise level for 100 hidden spectra case, around 2 for 50 hidden spectra. Below this limit, none of the method is able to detect the CH$_{4}$ spectra from the noise. For 20 hidden spectra, even at a factor above noise level of 3, none of the methods is able to detect the CH$_{4}$ spectra. One can also note that the psNMF is less stable since the standard deviation is much larger.

\section{Simulation of NOMAD-SO}

\subsection{Simulation dataset}

This second dataset has been generated with the most precise direct model,
taking into account the full non-linear radiative transfer and instrumental effects to
produce synthetic transmittance, highly comparable with actual observations.
Synthetic transmittances were made for real NOMAD-SO observation files using the relevant geometry and instrument parameters to attempt to include the variability inherent in the true measurements.

Model atmospheres for each occultation were developed from the GEM-Mars general circulation model \citep{Neary2018, Daerden2019}. 
The output of the model were provided for 1 Martian day every 10 solar longitude, and 48 timesteps per Martian day. 
Atmospheric profiles were developed for each occultation by interpolating the model temperature and pressure to the solar longitude, local solar time, latitude, longitude, and tangent altitude relative to the areoid.

To construct the simulated transmittance spectra, the high resolution irradiances were computed for each occultation assuming a spherically symmetry and the tangent atmosphere developed from GEM-Mars for several different abundance of methane and water, which were simulated as constant volume mixing ratios. The spectroscopic data for methane and water were taken from HITRAN 2016 using CO$_{2}$ broadening \citep{Gordon2017, Gamache_spectrallinelist_JoMS2016, Fissiaux_CO2broadeningcoefficients_JoMS2014}. 
The instrument forward model was then applied to each simulation by considering the AOTF bandpass, instrument Instrument Line Shape (ILS), blaze function, spectel to wavenumber calibration, and the contribution of light coming from the main order and nearby orders. 
The final synthetic transmittance spectra is the ratio of this low-resolution irradiance to the top-of-atmosphere low resolution irradiance.

The AOTF/echelle instrument was modeled using the latest available calibration \citep{Liuzzi2019, aoki2019}, considering order addition from $+/-2$ nearby orders (5 total). 
The spectral calibration of NOMAD-SO varies because it is affected by the instrument temperature, and is provided for each individual NOMAD spectra.  The 320 spectels cover the range 3056.1 cm$^{-1}$ to 3080.4 cm$^{-1}$ with a wavenumber step of 0.0763 cm$^{-1}$.

No simulation of dust has been performed. Due to the limited spectral range on a single order, about 25 cm$^{-1}$, the major effect of dust and other aerosols is relatively flat baseline, which we remove at the pre-treatment of the spectra. When dust is optically thick, then non-linearity may appear that are out of the scope of this simulation.

The simulation dataset consist of 12486 spectra, simulating observations
of order 136 in the same configuration as the 106 solar occultations
actually observed from May to December 2018. 

We add to the dataset a random noise with standard deviation of 0.001
and 0.0001 in order to simulate the instrumental noise (corresponding to SNR of 100 and 1000 approximately). 

We hide spectra containing CH$_{4}$
in a fraction of the total number of spectra from 1\% to 100\% in a random manner. In real observation, CH$_{4}$ may be spatially / temporally coherent but the number of scenarios is infinite. We feel that the random case is interesting enough to be tested. One has to note that contrarily to the previous toy model of section \ref{sec:synthetic-tests}, here abundance are quantitative abundance in the atmosphere.

The simulation parameters
are summed up in table \ref{tab:simulation_parameters}.

\begin{table}
\begin{tabular}{|c|c|c|c|c|}
\hline 
 & CH$_{4}$ {[}ppt{]} & H$_{2}$O {[}ppm{]} & fraction of CH$_{4}${[}\%{]}& noise level\tabularnewline
\hline 
\hline 
Value & 0; 100; 500; 1000 & 0; 10; 100 & 1; 5; 10; 50; 100 & 0.001; 0.0001\tabularnewline
\hline 
\end{tabular}

\caption{Simulation parameters. Fraction of CH$_{4}$ is fraction of spectra containing methane hidden in the simulation dataset. \label{tab:simulation_parameters}}
\end{table}

\subsection{Detection limits}

We applied the psNMF method with $N_{S}=5$, which is the most promising
one from the previous analysis. We compute the analysis 10 times for
10 different random noise realizations and average the results in order
to present robust conclusion. We select a pure CH$_{4}$ and a pure
H$_{2}$O spectra (noted $P_{CH_{4}}$ and $P_{H_{2}O}$) from the
simulation as reference spectra. 

\subsubsection{Methods to analyze the results}

 The main difference with the toy model section in \ref{sec:synthetic-tests} is that H$_{2}$O and CH$_{4}$ may be highly mixed in the sources. Simple correlation coefficient to pick the best source is thus not efficient enough. We propose here another approach to estimate the best source.
 
For each estimated source $\dot{S}_{:i}$, we analyze it as a linear
mixture of $P_{H_{2}O}$ and $P_{CH_{4}}$:

\begin{equation}
\dot{S}_{:i}=P_{H_{2}O}.\alpha_{H_{2}O,i}+P_{CH_{4}}.\alpha_{CH_{4},i}\label{eq:source_supervised_unmixing}
\end{equation}

This problem is called supervised detection algorithm since $P_{H_{2}O}$
and $P_{CH_{4}}$ are known, contrary to the general one, presented in Eq. \ref{eq:LinearMixture},
where source spectra are not known. The source $i^{*}$ with the
maximum $\alpha_{CH_{4},i^{*}}$ is selected as the best target CH$_{4}$ source,
called \textit{best source} hereafter.

We then propose to use three indicators of good detection :

\begin{itemize}
\item \emph{Fraction of the 4 main CH$_{4}$ peaks detected} (at 3057.7, 3063.4, 3067.2 and 3076.6 cm$^{-1}$). This is computed
using the peak detection algorithm from \textcopyright Matlab on both simulation and best source
with a tolerance of 2 spectels, i.e. detected peaks can be 2 spectels off
the expected one. The peak must be with a maximum amplitude larger than 1/1000 the maximum of $\dot{S}_{:i^{*}}$ to be considered significant. Please note that even there are only 5 possible fraction (0, 0.25, 0.5, 0.75 and 1), since we average on 10 realizations, any number can appear.

\item \emph{Mean distance to the expected center}. Mean distance in spectel between the CH$_{4}$ peaks detected in the
best source and the reference one.

\item \emph{Abundance of CH$_{4}$ in the source}. $\alpha_{CH_{4}}$ (from Eq. \ref{eq:source_supervised_unmixing}), which describes the amplitude
of the CH$_{4}$ peaks in the best source.
\end{itemize}

\subsubsection{Analysis of the results}

\begin{figure}


\hfill{}\includegraphics[width=1.0\columnwidth]{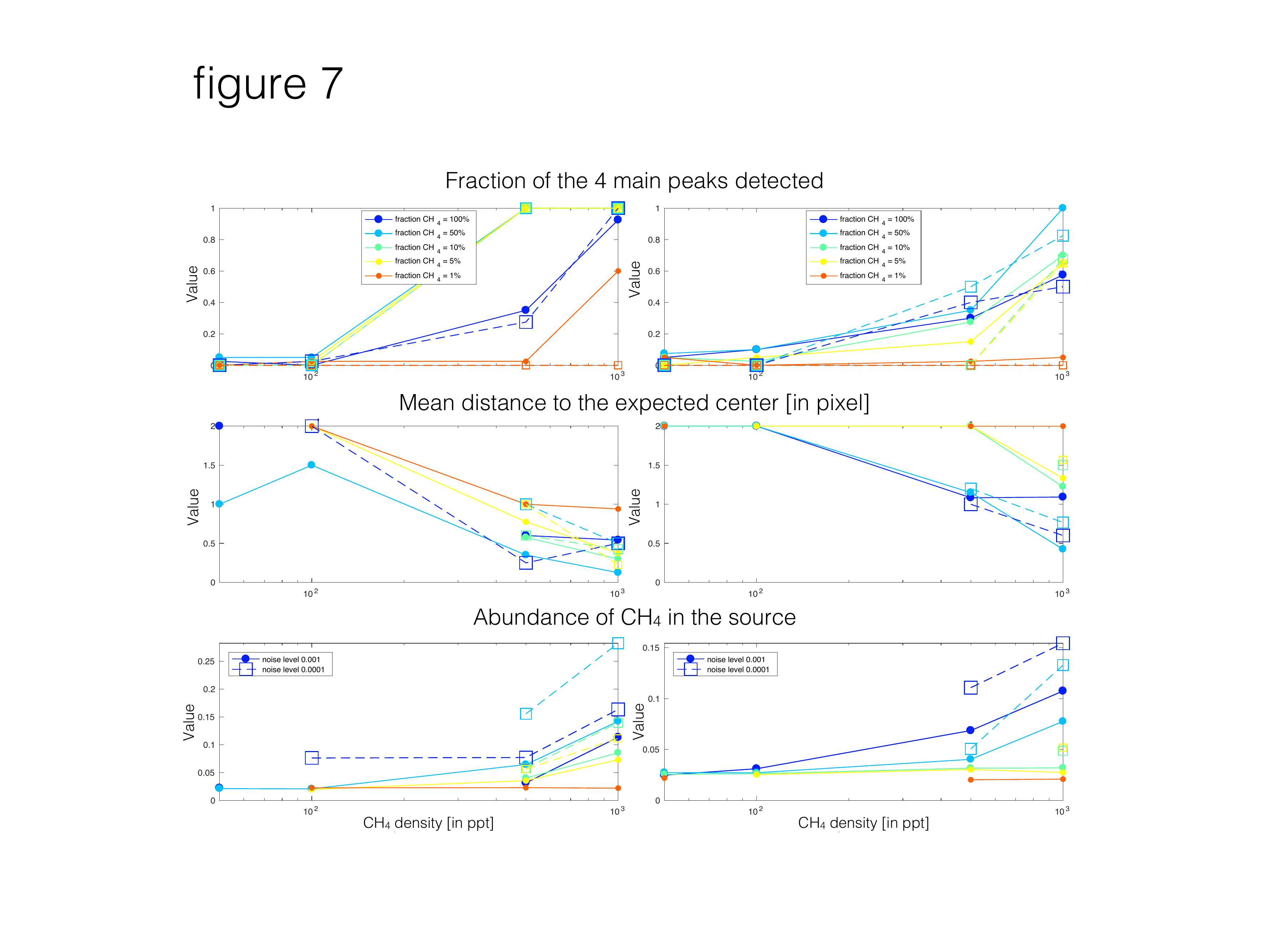}\hfill{}

\caption{Results of the psNMF algorithm for $N_{S}=5$ on simulation dataset,
averaged over 10 noise realizations, for different noise levels (0.001
and 0.0001) and different fractions of hidden CH$_{4}$ (1\%, 5\%,
10\%, \%, 100\%). Hidden CH$_{4}$ are taken within the same orbital
sequences. The left panels represent results for 10 ppm of water vapor and the right ones for 100 ppm of H$_2$O. From top to bottom, we show: a) Fraction of the 4 main CH$_{4}$ peaks detected in the \textit{best source} ; b) Mean distance to the expected center in spectel and c) Abundance of CH$_{4}$ in the source $\alpha_{CH_{4}}$. Please note that the absence of plotted data means that no source was successfully detected. \label{fig:results_simulation}}
\end{figure}

Figure \ref{fig:results_simulation} summarizes all the results. Fraction of the 4 main CH$_{4}$ peaks detected in the most relevant source has always a standard deviation $<$ 0.43 and a mean value of 0.06 over the 10 realizations. The Mean distance to the expected center has always a standard deviation $<$ 0.40 and a mean value of 0.07 over the 10 realizations. The abundance of CH$_{4}$ in the source has always a standard deviation $<$ 0.05 and a mean value of 0.005 over the 10 realizations. 

This figure shows that the detection limits
clearly depend on CH$_{4}$ density, but also on the fraction of hidden
CH$_{4}$ and noise level, as expected. Abundance of CH$_{4}$ in the source $\alpha_{CH_{4}}$ maximum is 25\%,
meaning that in any cases H$_{2}$O is dominating the best source
and so both CH$_{4}$ and H$_{2}$O are present in each best source.
This is because CH$_{4}$ is a minor specie (as expected from the conditions of our simulation), its absorption band generally follows the air-mass, as H$_{2}$O does.
So there is no particular source for CH$_{4}$ only.

When more than two lines are detected, we can consider it as a detection. This limit is reached for CH$_{4}\geq$ 500 ppt for 10 and 100 ppm of H$_{2}$O.
Nevertheless, the detection limits lies between 100 and 500 ppt in the case of 10 ppm of H$_{2}$O vapor since the detection is perfect (100\% of the 4 main CH$_{4}$ peaks detected) occurs for a fraction of CH$_{4}$ 5 to 50\%. Interestingly, the optimum detection is not when 100\% of the spectra contains CH$_{4}$, but more between 5-50 \%. This behavior is due to the statistics that is richer when also CH$_{4}$ is lacking in certain spectra. When 100\% of spectra contain CH$_{4}$, the statistical variability of the dataset is mainly due to airmass (atmosphere is assumed to be well mixed). So both CH$_{4}$ and H$_{2}$O are varying together and there is less statistics to base the detection on.

Noise level does not affect first the fraction of the 4 main CH$_{4}$ peaks but increases the spectral shift of the band center. In addition, it clearly affects the abundance and thus the band depth.

In conclusion, from this simulation analysis, one could expect detection limits of CH$_{4}$ in the range 100-500 ppt when operating in favorable conditions.

\section{Real data analysis}

\begin{figure}

\hfill{}\includegraphics[width=1.0\columnwidth]{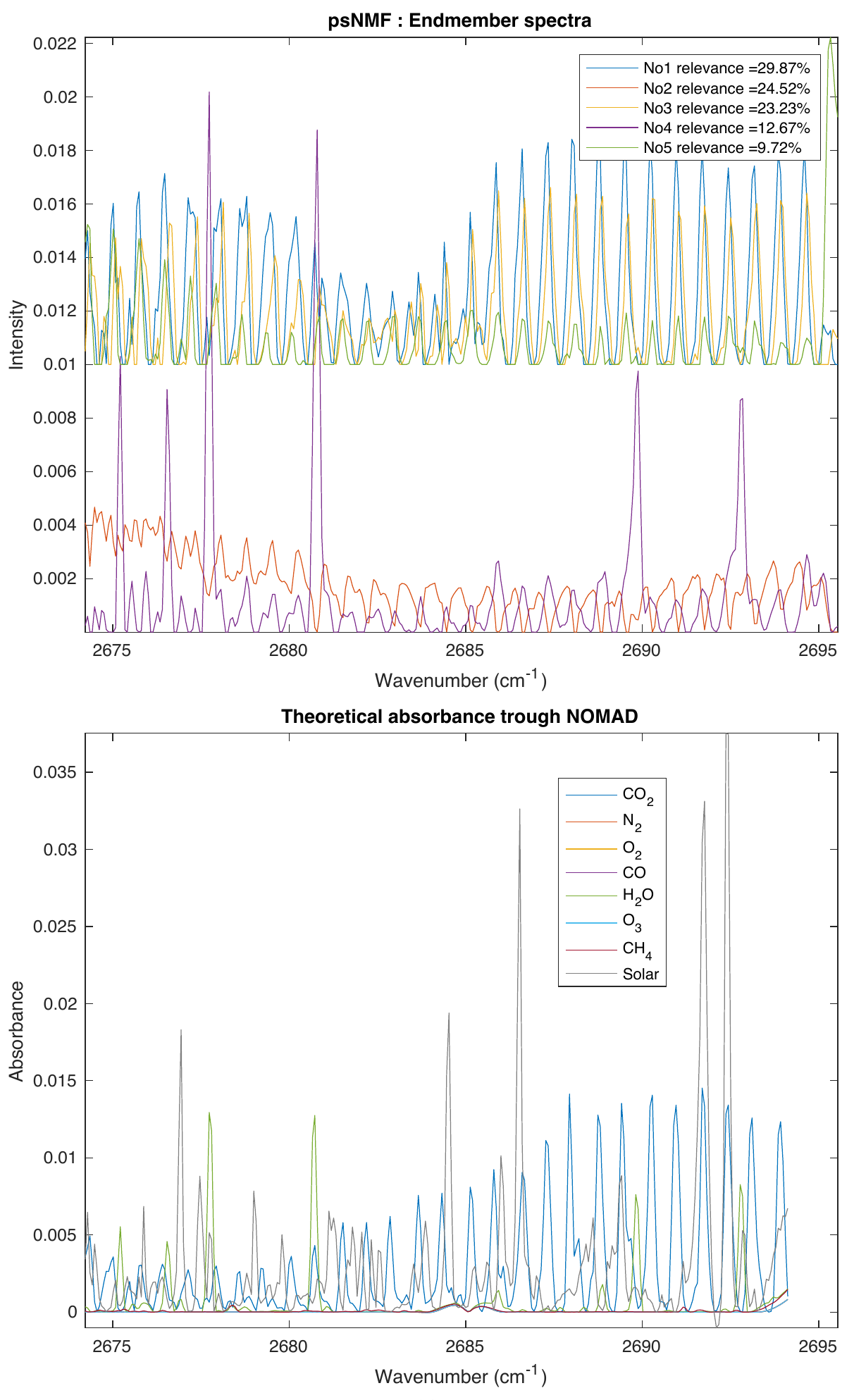}\hfill{}

\caption{Results of the psNMF algorithm for the diffraction order 119 for $N_{S}=5$. The sources
1, 3 and 5 are identified to CO$_{2}$ (shift of 0.01 for clarity). The source 2 is identified
to the background level (continuum misestimation). The source 4 is identified
to H$_{2}$O. No source seems to be related to CH$_{4}$. \label{fig:Results_real_order_119}}
\end{figure}

\begin{figure}

\hfill{}\includegraphics[width=1.0\columnwidth]{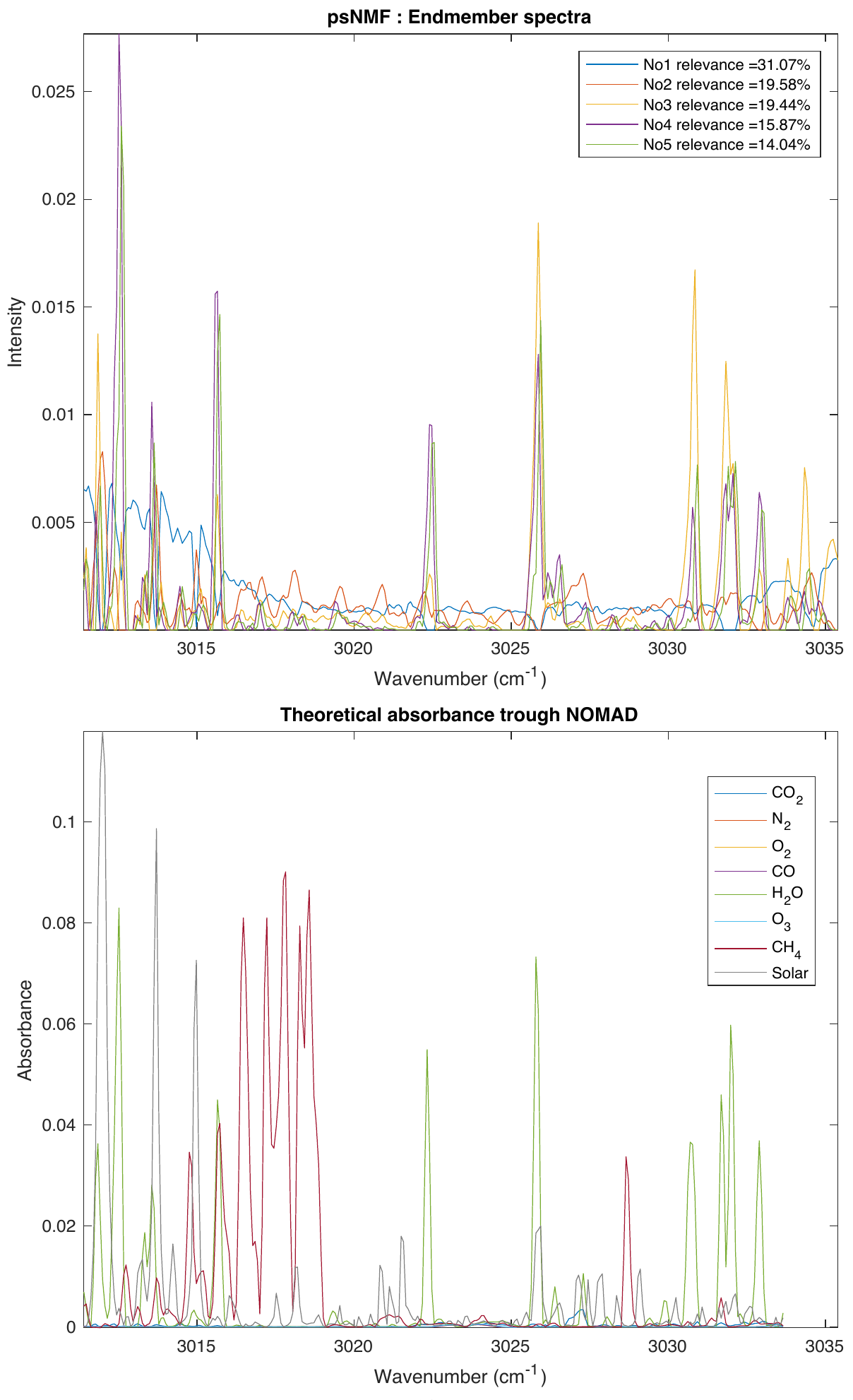}\hfill{}

\caption{Results of the psNMF algorithm for the diffraction order 134 for $N_{S}=5$. The source
1 is identified to the background level (continuum misestimation), the sources
3, 4 and 5 are identified to H$_{2}$O. The sources 2 present unmodeled
lines that are not present in the spectroscopic database. These lines has been first detected in the ACS instrument data and attributed to CO$_{2}$ magnetic dipole transition \citep{Trokhimovskiy_AA}. No source
seems to be related to CH$_{4}$.
 \label{fig:Results_real_order_134}}
\end{figure}

\begin{figure}

\hfill{}\includegraphics[width=1.0\columnwidth]{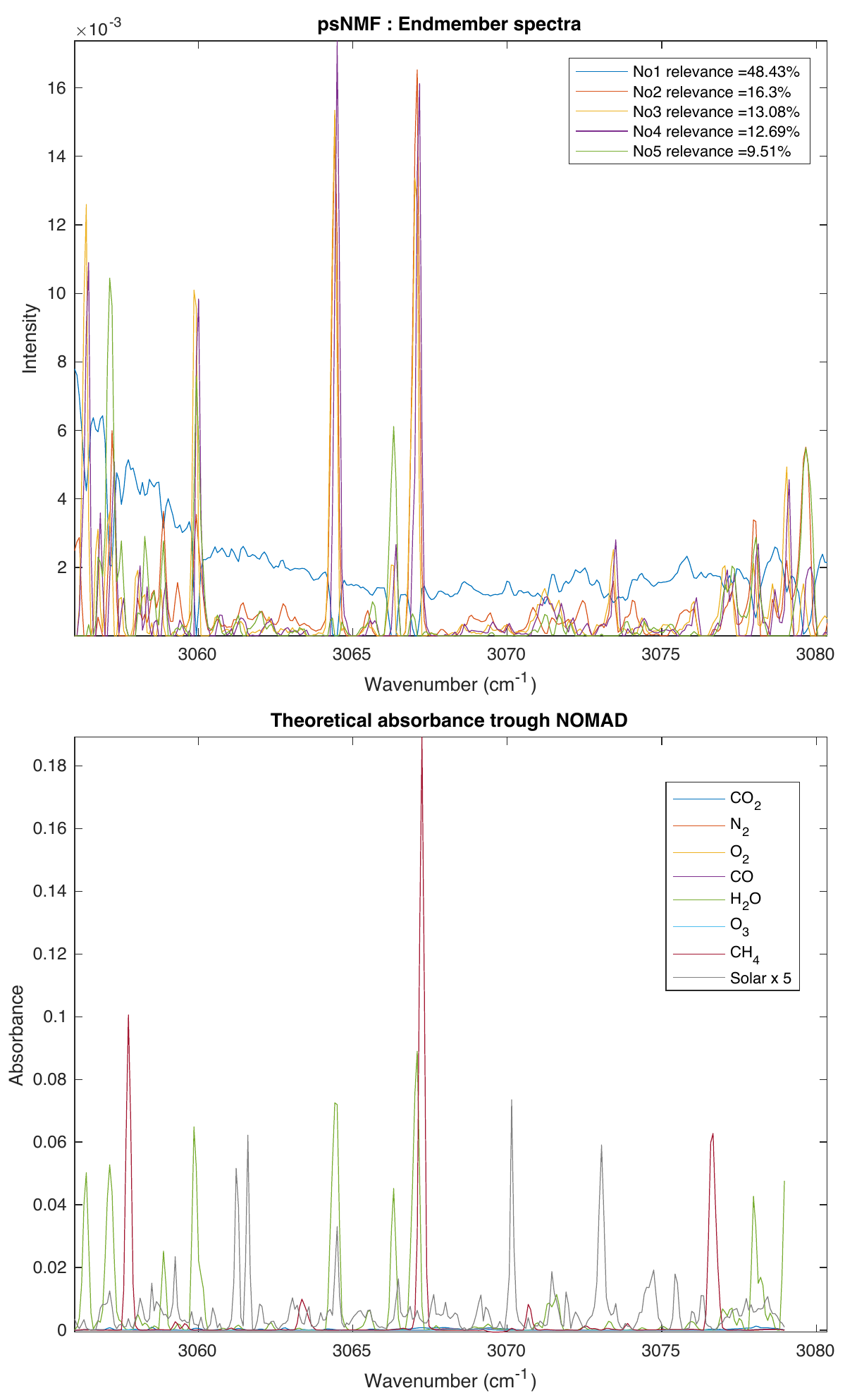}\hfill{}

\caption{Results of the psNMF algorithm for the order 136 for $N_{S}=5$. The source
1 is identified to the level background (continuum misestimation), the sources
2, 3, 4 and 5 are identified to H$_{2}$O, either directly either from
the adjacent orders. No source seems to be related to CH$_{4}$.
\label{fig:Results_real_order_136}}
\end{figure}

In this section, we report the results of actual NOMAD data, focusing
on diffraction orders with potential CH$_{4}$ lines: 119, 134 and 136, are shown
respectively on Fig. \ref{fig:Results_real_order_119}, \ref{fig:Results_real_order_134}
and \ref{fig:Results_real_order_136}. We used the 821 ingress and egress transit orbits for order 119, 2358 orbits for order 134 and 703 for order 136. We filter spectra with SNR $>$ 100. Results are compared with NOMAD
simulations \citep{Villanueva_PlanetarySpectrumGenerator_JoQSaRT2018}
using the calibration pipeline. This process adds ghost lines
from adjacent orders, as in real data. Table \ref{tab:Results_real_NOMAD}
summarizes the relative error and the number of spectra. The approach here is to compute the analysis with psNMF using $N_{S}=5$ in agreement with the previous section.

Please remind that our approach is fully blind: no spectral information has been included in the analysis (nothing about H$_{2}$O, CO$_{2}$ or CH$_{4}$).

For all orders, sources of H$_{2}$O are estimated, as expected.
Also a source presenting a residual of the continuum is always
present. Due to non-linearities of the radiative transfer, the acquisition
process (temperature dependence) and the wavenumber shift, the molecular species appears sometimes in different sources. 

Order 136 gives the 1 source related to the background and 4 sources related to H$_{2}$O. All 4 sources of water have the peaks but with different relative intensities and wavenumber shift.

For order 119, both CO$_{2}$ and H$_{2}$O lines are identified (see
Fig. \ref{fig:Results_real_order_119}). Since those two components
are uncorrelated, separated sources are found by the algorithm.

Interestingly, order 134 presents a source with unexpected lines. The main lines are at positions : 3016.70, 3017.07, 3018.12, 3019.54, 3020.90, 3022.25, 3023.60, 3024.96, and 3027.29 cm$^{-1}$. These lines has been also detected in the ACS instrument data and attributed to CO$_{2}$ magnetic dipole transition \citep{Trokhimovskiy_AA}. Further analysis shall be done to compare both NOMAD AND ACS data. 

Solar lines are never appearing in the sources. They are self-corrected by the calibration since we don't use a reference solar spectra but the solar observation during the transit when the tangent altitude is so high that there is no martian atmosphere (typically $>$ 200 km).

None of the analyzed orders presents sources related to CH$_{4}$.

\begin{table}
\begin{tabular}{|c|c|c|c|}
\hline 
 & 119 & 134 & 136\tabularnewline
\hline 
\hline 
$N_{O}$ & 134045 & 365985 & 140064\tabularnewline
\hline 
$N_{S}=4$ & 0.476 & 0.575 & 0.634\tabularnewline
\hline 
$N_{S}=5$ & 0.456 & 0.553 & 0.609\tabularnewline
\hline 
$N_{S}=6$ & 0.442 & 0.553 & 0.585\tabularnewline
\hline 
$N_{S}=10$ & 0.410 &  0.484 & 0.544 \tabularnewline
\hline 
\end{tabular}

\caption{Number of spectra $N_{O}$ and $RMSD$ relative errors for 4 to 10 number of sources $N_{S}$  resulting from the analysis of all observations of NOMAD data up to 15
January 2020, using the psNMF algorithm. $RMSD$ is computed from Eq.
\ref{eq:RMSD}. \label{tab:Results_real_NOMAD}}
\end{table}

\section{Discussions and Conclusion}

We implemented a new strategy to analyze spectroscopic datasets. 
This strategy is fully unsupervised, so that any kind of absorption
bands can be discovered. The amount of prior information required is thus very low. The computation can be done on a regular hardware for the most common database and within reasonable amount of time ($\sim$100000 spectra). 

We illustrate the approach for typical atmospheric spectroscopy. We first put forward a synthetic test, based on simple linear mixing to give a toy example and 
to identify the best promising algorithm. The psNMF clearly outperformed MU and BPSS2.

Then we proposed a simulation, based on realistic radiative transfer and instrumental effects, applied on NOMAD-SO spectra. 
The detection limits goes below 500 ppt in favorable conditions, with reduced H$_{2}$O and low noise level.  The same range of detection limits is reach with usual approach of model fitting at a much higher computation cost and analysis effort. Given the simplicity of use, this tool may be relevant to handle large and complex datasets at first glance. As a perspective, analysis of residuals after the non-linear retrieval of the data may lower the detection limits. One can then test if the residuals are simply Gaussian noise, or if they may contain interesting features.

Interestingly, a molecular specie not well mixed in the atmosphere can be most easily detected with our approach.

The last section presented the results of the application on real NOMAD-SO data, using orders 119, 134 and 136, 
selected as they are representative of the baseline strategy of measurements in NOMAD, allowing characterization of H$_2$O and potential detection of CH$_4$.
The outcome is that no CH$_{4}$ has been identified, but H$_{2}$O and CO$_{2}$ are detected. 
Interestingly a new set of spectral lines has been discovered in the NOMAD data. These lines has been first detected in the ACS instrument data and attributed to CO$_{2}$ magnetic dipole transition \citep{Trokhimovskiy_AA}. We thus confirm their presence with our current analysis.

One way to go back to the data is to pick the real data with the highest source contribution $\dot{\mathbf{A}}$. Our quicklook analysis is thus only a starting point of a more complete scientific analysis. This second step will require much more prior information (chemical compounds, fundamental spectroscopic constants, radiative transfer model, ...).

Future work should apply the proposed approach to other datasets, such as other NOMAD-SO orders, or other spectroscopic datasets (including hyperspectral images)
from laboratory measurements, ground based telescopes or space-born spectrometers. The approach is generic enough to treat datasets that can be at first order approximated to a linear mixture.

\subsubsection*{Acknowledgements}

We acknowledge support from the ``Institut National des Sciences
de l'Univers'' (INSU), the \textquotedbl Centre National de la Recherche Scientifique\textquotedbl{} (CNRS) and \textquotedbl Centre National d'Etudes Spatiales\textquotedbl{} (CNES) through the \textquotedbl Programme National de Plan\'etologie\textquotedbl{} and the ExoMars TGO programs. The NOMAD experiment is led by the Royal Belgian Institute for Space Aeronomy (BIRA-IASB), assisted by Co-PI teams from Spain (IAA-CSIC), Italy (INAF-IAPS), and the United Kingdom (Open University). This project acknowledges funding by the Belgian Science Policy Office (BELSPO), with the financial and contractual coordination by the ESA Prodex Office (PEA 4000103401, 4000121493), by Spanish Ministry of Science and Innovation (MCIU) and by European funds under grants PGC2018-101836-B-I00 and ESP2017-87143-R (MINECO/FEDER), as well as by UK Space Agency through grants ST/R005761/1, ST/P001262/1, ST/R001405/1 and ST/R001405/1 and Italian Space Agency through grant 2018-2-HH.0. This work was supported by the Belgian Fonds de la Recherche Scientifique - FNRS under grant number 30442502 (ET-HOME). The IAA/CSIC team acknowledges financial support from the State Agency for Research of the Spanish MCIU through the Center of Excellence Severo Ochoa award for the Instituto de Astrof{\'i}sica de Andaluc{\'i}a (SEV-2017-0709). US investigators were supported by the National Aeronautics and Space Administration. Canadian investigators were supported by the Canadian Space Agency.

\bibliographystyle{elsarticle-harv}

\end{document}